\documentclass[conference,a4paper]{IEEEtran}
\usepackage{graphicx}
\author{Gitanjali Bhutani\\Alcatel-Lucent Technologies India\\ bhutanig@alcatel-lucent.com }
\title{A Round-based Pricing Scheme for Maximizing Service Provider's Revenue in P2PTV Networks}
\date{}
\newcommand{\tablefontsize}{\small}
\begin{document}
\maketitle
\begin{abstract}
  In this paper, we analyze a round-based pricing scheme that encourages favorable behavior from users of real-time P2P applications like P2PTV. In the design of pricing schemes, we consider price to be a function of usage and capacity of download/upload streams, and quality of content served. Users are consumers and servers at the same time in such networks, and often exhibit behavior that is unfavorable towards maximization of social benefits. Traditionally, network designers have overcome this difficulty by building-in traffic latencies. However, using simulations, we show that appropriate pricing schemes and usage terms can enable designers to limit required traffic latencies, and be able to earn nearly 30\% extra revenue from providing P2PTV services. The service provider adjusts the prices of individual programs incrementally within rounds, while making relatively large-scale adjustments at the end of each round. Through simulations, we show that it is most beneficial for the service provider to carry out 5 such rounds of price adjustments for maximizing his average profit and minimizing the associated standard deviation at the same time. 
\end{abstract}
\section{Introduction}
\label{sec:introduction}
The past few years have witnessed a significant growth in the usage and utility of P2P networks. These networks have grown from earlier basic ones like Napster and Gnutella to more recent and sophisticated ones like BitTorrent. P2P networks provide an effective mechanism for resource sharing, allowing sharing of memory and processing power. This allows workload distribution and load balancing making computing systems more efficient. P2P networks also provide advantages such as scalability and reliability that are important characteristics of distributed systems. They make information sharing easy and make information pools easily accessible to a large group of users. 

P2P networks originally emerged to provide extensive file and information sharing. These systems undoubtedly met their requirements and introduced a new trend in information sharing making it easier and widely accessible to a large group of users. With the emergence of more sophisticated P2P networks, their capabilities were used to add a new dimension to resource sharing. Systems on these networks can now share resources like storage space and processing power, thus providing a higher performance at lower costs. A related advantage is workload distribution enabling parallel processing among systems on the network, thereby leading to a more efficient, robust and reliable computing system.

The lack of central administration in P2P networks ensures that there is no need for users to trust a single authority or depend on one for their operation. An emerging trend is the harnessing of advantages of P2P networks in the television domain, commonly called P2PTV. Software applications supporting P2PTV are designed to redistribute video streams or files on a P2P network. They allow users to watch the content as it is being downloaded, and each user, while downloading the content also uploads it to other interested peers thereby contributing to the overall available bandwidth. Such behavior provides significant savings in bandwidth while also making downloading and sharing of live content cost-effective and simpler. By allowing content to be downloaded by peers from peers, P2PTV networks reduce the load on the central content generating server. Due to the server workload reduction and the ability to share TV channels among peers, P2PTV makes TV channels available to a larger community of users at reasonable costs.

However, the design of P2PTV systems must take into account selfish nodes and their refusal to cooperate and share content. This paper discusses a round-based pricing adjustment scheme to provide incentives to users as the TV channel content is downloaded and viewed by increasing number of users. It explores a mechanism that maximizes the revenue of the P2PTV service provider while providing incentives to users and encouraging them to share content and download from a closely situated peer.

This paper is organized as follows: Section~\ref{sec:literature-survey} reviews the work that has been carried out in scientific literature in the area of P2P networks and pricing in general. We argue that the problem of pricing in P2PTV networks is different from the problem of pricing in P2P networks, which has been handled before in the earlier literature. Section~\ref{sec:experimental-setup} describes the setup of the simulation experiments, and details various assumptions and models we have used for modeling the network, the users, their content requests and pricing schemes. In section~\ref{sec:results}, we report the results obtained through such experimentation, and conclude the paper with a discussion on results and future scope of work in section~\ref{sec:conclusions}.

\section{Literature Survey}
\label{sec:literature-survey}
The literature pertaining to P2PTV can be divided into three sub-categories. One part of the literature deals with application level multicast and effective video streaming to be able to provide P2PTV users with high-quality video service. The reader is referred to \cite{Banerjee2002, Banerjee2004, Chu2000, Dominguez2007} for a detailed description of the work carried out in this area.

Second part of the literature deals with design of P2P networks to support real-time technologies like P2PTV, whereby unicast networks can be effectively replaced with P2P-based networks with as little a bandwidth usage overhead as possible. However, in most studies it is found that many redundancies need to be built in to the network design for P2P networks to be able to replicate the QoS requirements of a real-time application like P2PTV (\cite{Cohen2002, Lam2004, Zhao2004}). Some works also compare alternative strategies of constructing redundancies for P2P networks (\cite{Rodrigues2005, Weatherspoon2002}).

P2P networks heavily rely on their user penetration and participation in networks for realizing their promised benefits. This third part of the literature related to user behavior in P2P networks is itself divided into three sub-parts. One large sub-part talks of selfish peers who significantly undermine the advantages of P2P networks, and deals with the problem of free-riders. Studies show that as much as 70\% of users in a single Gnutella network are free-riders. (\cite{Feldman2004, Ma2004, Wierzbicki2005, Wongrujira2005a}) talk of various incentivization schemes to overcome this and to encourage users to share resources. Various incentive schemes have been proposed to encourage user cooperation in P2P systems \cite{Feldman2005a} - Inherent generosity (\cite{Camerer2003, Feldman2004}), Monetary Payment Schemes \cite{Golle2001} and Reciprocity based schemes (\cite{Andrade2005, Cohen2003}). 

The second sub-part of the user behavior related part of literature deals with the problem of non-cooperation of peers in forwarding lookup messages. In \cite{Huang2006a}, a scheme is setup to allow all lookup-message forwarders to share profits. In \cite{Gupta2004}, forwarders get monetary incentives and schemes are established to prevent peers from overcharging.

The third sub-part of literature deals with the problem of topology in P2P networks. Peers exploit the principle of locality and usually attempt to minimize costs by selecting peers that are closest to them. However, such behavior adversely impacts the topology of P2P networks \cite{Moscibroda2006}. This sub-part of the literature also studies various algorithms for selection of peers to optimize performance of P2P networks. 

In addition to the work carried out in the networking-related literature, we also look at the work carried out in the area of revenue management and pricing. The reader is referred to \cite{Phillips2005,Talluri2004} for a comprehensive treatment on pricing and revenue optimization models. In addition, \cite{Bitran2003,Elmaghraby2003} present a broad overview of pricing and revenue management issues and provide a unified modeling framework focused on dynamic pricing models in revenue management. \cite{Gallego1997,Kleywegt2001,Maglaras2004} formulate complex stochastic dynamic programming models for the pricing problem, where demand is modeled as a stochastic process with price-dependent intensity. These works provide structural results and heuristics based on the deterministic versions of the pricing problem. In particular, \cite{Paschalidis2000,Yeung2006,Zghaibeh2006} look at the pricing problem in context of P2P networks.

This paper looks at the problem of pricing in P2PTV networks, especially with the aim of encouraging socially-correct behavior amongst users. Note that as the content and network bandwidth are both priced, the service provider should be able to incentivize certain uses of the network bandwidth to induce desired behavior amongst users. The utility of the content served in P2PTV networks for its users can be expected to be time-sensitive. As the scheduled timing of the program content being offered passes by, the utility for the users would fall down significantly, and it is in this respect that the content can be characterized more or less as perishable in nature. It is due to this perishable nature of the content on offer that we contend that the problem of pricing and inducing socially correct behavior in P2PTV networks should differ from that in P2P networks. Towards achieving this aim, in this paper we rely on a simulation-based modeling and experimentation to estimate the impact of a pricing and incentivization based scheme on profits earned by the service provider. 

\section{Experimental Setup}
\label{sec:experimental-setup}
\subsection{Modeling the Network}
\label{sec:modeling-network}
For our experimental setup, there are two main characteristics that would play an important role in deciding the benefits users and the service provider observes from offering P2PTV on a particular network. One characteristic is the density and the interconnectivity existing in the network, and the other is the link costing prevailing in the network. Networks that are denser in structure than others are likely to provide their users with choice of connecting to a variety of peers, and can be expected to serve as a better ``breeding ground'' for Peer-to-Peer applications like P2PTV. In our work, we estimate the density of a network through the ratio of the number of interconnections to the number of users. Link costing in viewing a P2PTV program has important connotations in terms of the demand the network would see for the P2PTV programs being served on it. Larger link costs would serve as a deterrent for users in subscribing to various programs, as many of them would find that the link costs plus content costs of the programs to be larger than their willingness-to-pay. In such cases, such users would altogether abstain from watching any P2PTV program. Link costs can have a variety of cost components and pricing schemes - some of them are the setup costs, operational or maintenance costs, pay-per-usage costs and pay-for-connected-time costs. Some networks like an Ethernet-based LAN would have high setup costs but low operational and pay-per-usage costs, while others like Dialup-based connections would have low setup costs but high pay-per-usage and pay-per-connected-time costs. In our simulations, we assume that the link cost between two nodes of the network is a function of the distance between them. In real-life the function can be expected to have a step-wise linear form similar to the one shown in Figure 1. Note that users have a choice in the type of network they use to connect to their peers even for a given physical distance between them. For shorter distances, networks like WiFi and Ethernet-based LANs may be used, while for larger distances Dialup and DSL based connections may be used. 

\begin{figure}
  \includegraphics[width=240pt]{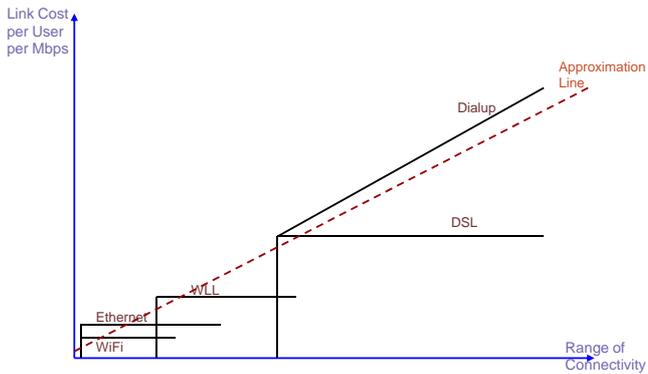}
  \caption{Characteristic Link Costs versus amount of data transferred for various types of networks and representative Approximation Line}
  \label{fig:characteristic-link-cost}
\end{figure}

Figure~\ref{fig:characteristic-link-cost} can be explained as follows. For networks with high setup costs and low pay-per-usage costs, the link cost can be assumed to a constant number that is representative of the amortized cost of installing and operating that network. Once peers are on such networks, the link cost does not depend upon the distance between them and is denoted by a constant value function like that depicted for Ethernet-based LANs. However, for networks with low setup costs and high pay-per-usage and pay-for-connected-time costs, the characteristics are different. For example, for Dialup based networks, the user pays for the length of time he is connected to the network. In such cases, as the distance between peers increases, the speed of connectivity between them reduces and the transfer of same amount of traffic takes longer thereby making link costs vary with time. This is shown in Figure~\ref{fig:characteristic-link-cost} for the link cost function of Dialup connections. However, with many network connectivity options being available for a given pair of peers separated by some distance, and with different characteristics of each of these connectivity options, the combined overlapping behavior is complicated and makes it difficult for us to comprehend the implications of our results. Hence, we approximate the combined overlapping link cost function of various networks by a straight line as shown in Figure~\ref{fig:characteristic-link-cost}.

\subsection{Modeling the Users}
\label{sec:modeling-users}
While simulating users, we assume that their willingness-to-pay is derived from a common, yet hidden mathematical model. In usual circumstances where users are not incentivized to reveal their willingness-to-pay (or utility from watching a P2PTV program), the service provider tries to estimate this hidden model of true demand, and prices the service accordingly. Such estimation is typically carried out over a period of time, where the service provider learns the right pricing through trial and error in the marketplace. Note that such experimentation with pricing in the marketplace through trial and error is costly, as it loses the service provider important revenue opportunities. 

\begin{figure}
  \includegraphics[width=240pt]{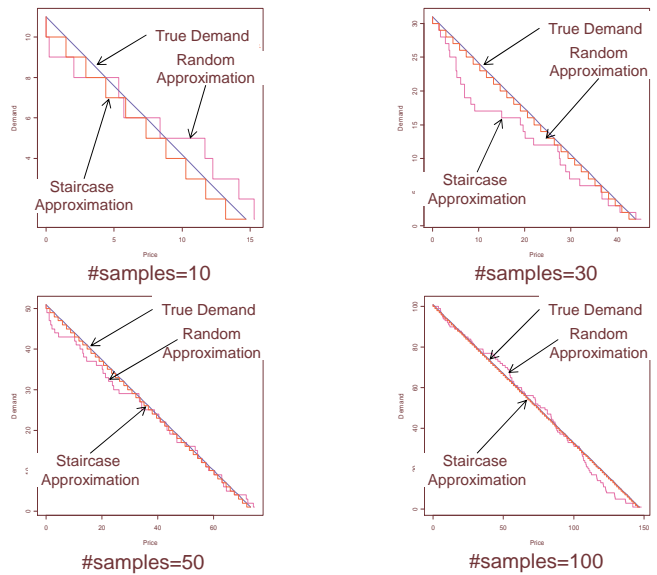}
  \caption{Estimation of true demand function through stair-case approximation and random draws for a single-dimensional demand}
  \label{fig:single-dim-demand-estimation}
\end{figure}

In a particular simulation run of the model, we can assume that the willingness-to-pay for each of the users is either systematically drawn from the underlying hidden model of true demand, or is a random draw from it. A set of systematically drawn willingness-to-pay numbers would look like the staircase approximation shown in Figure~\ref{fig:single-dim-demand-estimation}. Note that such systematic draws are based on step functions indicating the prices at which the demand increases by one unit as per the true demand function. On the other hand, the set of random draws would have many spikes in them, and would look similar to the underlying true demand model only when the number of users, and hence the number of draws is large in the statistical sense. The shape of the observed willingness-to-pay demand function with random draws with 10, 30, 50 and 100 draws is shown in Figure~\ref{fig:single-dim-demand-estimation}.  

\begin{figure}
  \includegraphics[width=240pt]{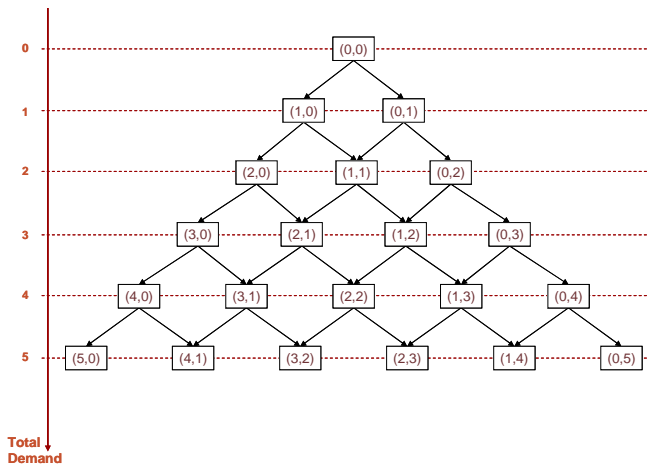}
  \caption{Computational Complexity associated with discrete approximation}
  \label{fig:comp-complexity}
\end{figure}

Note that while the stair-case approximation correctly depicts the true underlying demand function, and would allow us to compute the true revenue generated, it increases the computational complexity in calculating the various user scenarios, as depicted in Figure~\ref{fig:comp-complexity}. Figure~\ref{fig:single-dim-demand-estimation} denotes the error generated in estimating a one-dimensional demand function with stair-case approximation and random estimation. Note that the number of dimensions indicates the number of alternative programs that are available for the users to view at a particular instant of time. Figure~\ref{fig:comp-complexity} denotes the number of cases that need to be evaluated in generating the stair-case approximation of a two-dimensional demand function.  

\subsection{Modeling Content Requests}
\label{sec:model-cont-requ}
In this section, we develop the model for characterizing and simulating the content requests. Let us assume that there are $n$ programs being telecast simultaneously, and that each of the users makes the choice of watching one and at most one of these programs at a given time. A user's choice of program depends upon his utility from watching the program (thereby leading to his willingness-to-pay), and the network and the content prices he has to pay for watching the program. Note that in our models, the service provider does not do any differential pricing of the content to individual users. However, the network cost being paid by a user would typically depend on his P2P network behavior, with users depicting cooperative behavior being incentivized appropriately. The amount of incentives given to individual users would depend on the incentivization mechanism being adopted by the service provider.

Let us assume further that the maximum demand to be expected of each of these $n$ programs can be denoted by $D_1, D_2 \ldots D_n$ respectively. The service provider in practice can obtain these values by conducting market surveys. Let $d_1, d_2, \ldots d_n$ denote the actual demand for each of the programs when the content prices for each of them are $p_1, p_2, \ldots p_n$ respectively. Note that the amount of viewership for a program not only depends upon its own price, but also on the price of all other programs. This indicates that we can expect to have a certain degree of correlation and cross-dependencies between the demands for various programs. We use Equation~(\ref{eq:eq-1}) to model the overall demand function of the various programs offered by the service provider, where $\mathcal{D}$ denotes the demand vector for all programs, $\mathcal{P}$ denotes the current pricing vector set by the service provider and $\mathcal{H}$ denotes the matrix of self and cross-elasticities associated with the demand functions.
\begin{equation}
  \mathcal{D} = \mathcal{H} \times \mathcal{P}
  \label{eq:eq-1}
\end{equation}
The service provider strives to learn the unknown $\mathcal{H}$ by changing the pricing vector $\mathcal{P}$ and observing the resultant demand vector $\mathcal{D}$. For a given set of $M$ users in the system at a given point in time, following the random draws method of the previous section, the individual demands for each of the programs can be simulated as 
\begin{eqnarray}
  d_1 \sim U[0,\textrm{min}(M, D_1)]\\
  d_2 \sim U\left[0,\textrm{max}\left(0, \textrm{min}\left(M-d_1, D_2\right)\right)\right]\\
  \vdots\\
  d_n \sim U\left[0,\textrm{max}\left(0,\textrm{min}\left(M-\sum_{i=1}^{n-1}d_i, D_n\right)\right)\right]
\end{eqnarray}
where $U[a,b]$ denotes a random draw from a uniform distribution on the range $[a,b]$. As far as the simulation is concerned, $\mathcal{H}$ can be assumed to be known and the prices corresponding to a sample set of individual demands can be obtained by inverting Equation~(\ref{eq:eq-1}).

\subsection{Pricing Schemes and Learning Prices}
\label{sec:pric-schem-learn}
As a base case for our experiments, we consider that the users are offered incentives based on the number of programs they serve to their peers. Such a scheme prompts serving users to serve whole programs, rather than break in the middle. We compare the revenue of the service provider who provides such an incentive scheme with that of a service provider relying explicitly on unicast transmissions from its servers. Also, we choose an incentive scheme based on the number of programs served, rather than a revenue-sharing based incentive model.  Under a revenue-sharing model, there would be a tendency on the part of serving users to prefer serving high-value content to low-value content. Such a tendency would make serving the low-value content a high-cost venture for the service provider.

Incentive programs can also be characterized in terms of the times at which they require users to declare their intent. For our base case, we consider real-time sharing incentivization, where users declare their content requests in real-time, and the interested user servers declare their intent-to-share also in real-time. In contrast, in a know-ahead incentivization scheme, the users are the user servers declare their content requests and intents-to-share in advance. Such advance declaration of content requests and intents allows the service provider to exploit further revenue generating opportunities in the network. Any such revenue gains can be shared with users to create a win-win situation for both the users and the service provider. 

Price discovery under uncertain demand is a difficult problem to solve
\cite{Phillips2005, Talluri2004}. In this paper, we use a simple directional descent based learning mechanism for price discovery. The process generating the price discovery can be described as: Users arrive randomly at the service providers ``site'', and check the current prices of all programs slotted for a particular time-slot. They choose the program that gives them the largest difference between their willingness-to-pay and the price being requested. As soon as a user selects a particular program, the service provider compares its current demand with its target popularity and adjusts the prices accordingly. Also, as the users expect prices to change, they come back to the site at a later point in time to re-evaluate their decisions. Each time a user comes back to re-evaluate his decision before the start time of the programs is termed as a round. For the learning mechanism, the service provider makes small-step price corrections within rounds, and large-step corrections between rounds. 

\section{Results}
\label{sec:results}
\begin{figure}
  \includegraphics[width=240pt]{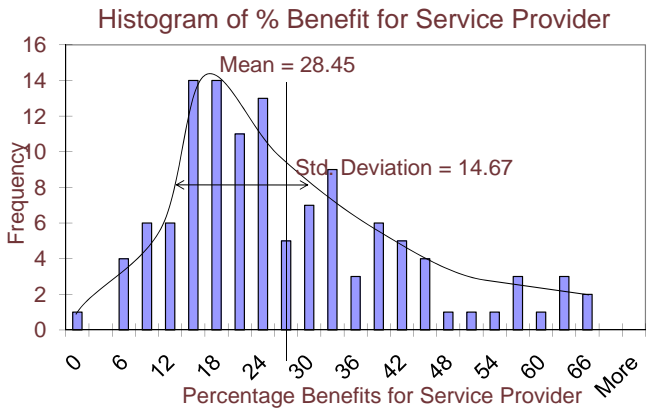}
  \caption{Percentage improvements in benefits in a P2PTV network over a unicast network}
  \label{fig:content-provider-benefits}
\end{figure}

For the setup of the simulation experiments described, we carried out our experiments with 60 users and 15 programs. We first assumed that the users would re-visit the service provider's site for pricing 2-3 times, and hence varied the number of rounds for our experiments between 2 and 3. The willingness-to-pay function of various users was simulated. We then calculated the revenues collected by the service provider for the cases where he runs his network as a P2PTV network, and find the percentage gains with respect to the revenues he would have generated by running it as a unicast network. Figure~\ref{fig:content-provider-benefits} plots the histogram denoting the percentage benefits accrued by the service provider by running his network as a P2PTV network, as against a unicast network, and following the base case incentive policy described in the previous section. 

\begin{figure}
  \includegraphics[width=240pt]{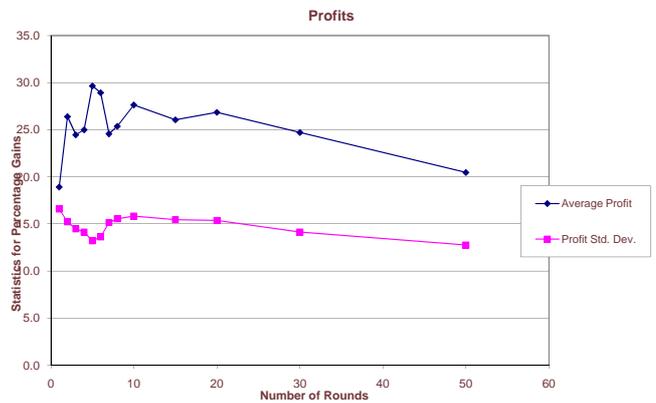}
  \caption{Service provider's average profit and associated standard deviation versus the number of rounds}
  \label{fig:avg-profit-std-dev-revisit}
\end{figure}

In a real-world setup, a service provider would try to influence user behavior in terms of the number of times they re-visit the site. We also looked at the potential revenue benefits the service provider can expect by influencing such behavior. Table~\ref{tab:table1} tabulates the average profit and the associated standard deviation observed by the service provider in P2PTV over and above the unicast network case when the number of rounds are varied. Figure~\ref{fig:avg-profit-std-dev-revisit} depicts these results in the form of a graph. We notice that for a value of number of rounds = 5, the service provider observes the largest average revenue gain, with the smallest deviation around it. 

\begin{table}[ht]
\centering
\caption{Service provider's average profit percentage over and above unicast network profits and associated standard deviations with variations in number of rounds}
\label{tab:table1}
\begin{tabular}{|r|c|c|}
\hline
\multicolumn{1}{|c|}{\textbf{\tablefontsize Rounds}} & \multicolumn{1}{|c|}{\textbf{\tablefontsize Average Profit}} & \multicolumn{1}{|c|}{\textbf{\tablefontsize Profit Std. Dev.}} \\
\hline
{\tablefontsize 1} & {\tablefontsize 18.9} & {\tablefontsize 16.6} \\
\hline
{\tablefontsize 2} & {\tablefontsize 26.4} & {\tablefontsize 15.2} \\
\hline
{\tablefontsize 3} & {\tablefontsize 24.5} & {\tablefontsize 14.5} \\
\hline
{\tablefontsize 4} & {\tablefontsize 25.0} & {\tablefontsize 14.1} \\
\hline
{\tablefontsize 5} & {\tablefontsize 29.7} & {\tablefontsize 13.2} \\
\hline
{\tablefontsize 6} & {\tablefontsize 28.9} & {\tablefontsize 13.6} \\
\hline
{\tablefontsize 7} & {\tablefontsize 24.6} & {\tablefontsize 15.1} \\
\hline
{\tablefontsize 8} & {\tablefontsize 25.4} & {\tablefontsize 15.5} \\
\hline
{\tablefontsize 10} & {\tablefontsize 27.6} & {\tablefontsize 15.8} \\
\hline
{\tablefontsize 15} & {\tablefontsize 26.1} & {\tablefontsize 15.4} \\
\hline
{\tablefontsize 20} & {\tablefontsize 26.8} & {\tablefontsize 15.4} \\
\hline
{\tablefontsize 30} & {\tablefontsize 24.7} & {\tablefontsize 14.1} \\
\hline
{\tablefontsize 50} & {\tablefontsize 20.5} & {\tablefontsize 12.8} \\
\hline
\end{tabular}
\end{table}

\section{Conclusions}
\label{sec:conclusions}
\subsection{Discussion on Results}
\label{sec:discussion-results}
Pricing schemes and incentivization policies are an important instrument in the hands of a P2PTV network designer. In this work, through use of simulations, we have proved that by use of suitable pricing schemes, the network designer can provoke users to display socially favorable behavior, and earn as much as 30\% more revenue on average than the case where he tries and builds a unicast network. 


In this paper, we have proposed a mechanism based on rounds of price adjustments, where the service provider adjusts prices of individual programs incrementally within rounds, while making large-scale adjustments only at the end of each of the rounds. We have experimentally found, as shown in Figure~\ref{fig:avg-profit-std-dev-revisit}, that it may be optimal for the service provider to have 5 such rounds of price adjustments to maximize his average profit and minimize the associated standard deviation at that same time.

Significant synergies exist between the business of a network service provider, and that of a P2PTV content-provider. Results of this paper suggest that companies in these individual businesses would do well by investing in the other business. In particular, since the network service provider business is more capital-intensive and larger entry barriers exist in trying to get into this business, we envisage that players from this business would start investing more and more into the P2PTV business.

\subsection{Scope for Future Work}
\label{sec:scope-future-work}
Networks have traditionally been modeled using static models that typically use the graph-theoretic approach of depicting users as Nodes of a Graph, and their interconnections using the Arcs of the Graph. In a P2P network, users often go out of the Network and come back up. The design of the network keeps changing in real-time, and an almost real-time optimization is required to continually get the maximum benefit out of P2P networks. Due to the dynamic nature of P2P networks, we argue that dynamic models are required to effectively model P2P networks and more work needs to be done in this area. However, as this is not the focus of the current work, we use a simplistic quasi-static model for modeling P2P networks and use it to derive our results.

In this work, we have used fairly simplistic models to characterize the network and the users. More detailed and accurate models can be used in further studies. 

The problem of profit maximization in view of user participation can be proved to be a non-linear optimization problem that can potentially have multiple local solutions. In this work, we have satisfied ourselves with local maximization results. These solutions do indicate the amount of benefit to be achieved by such an optimization routine. However, a global maximum would present the service provider with further opportunities to increase her profitability.

\bibliographystyle{IEEEtran}
\bibliography{literature/p2p}

\begin{thebibliography}{10}
\providecommand{\url}[1]{#1}
\csname url@samestyle\endcsname
\providecommand{\newblock}{\relax}
\providecommand{\bibinfo}[2]{#2}
\providecommand{\BIBentrySTDinterwordspacing}{\spaceskip=0pt\relax}
\providecommand{\BIBentryALTinterwordstretchfactor}{4}
\providecommand{\BIBentryALTinterwordspacing}{\spaceskip=\fontdimen2\font plus
\BIBentryALTinterwordstretchfactor\fontdimen3\font minus
  \fontdimen4\font\relax}
\providecommand{\BIBforeignlanguage}[2]{{%
\expandafter\ifx\csname l@#1\endcsname\relax
\typeout{** WARNING: IEEEtran.bst: No hyphenation pattern has been}%
\typeout{** loaded for the language `#1'. Using the pattern for}%
\typeout{** the default language instead.}%
\else
\language=\csname l@#1\endcsname
\fi
#2}}
\providecommand{\BIBdecl}{\relax}
\BIBdecl

\bibitem{Banerjee2002}
S.~Banerjee, B.~Bhattacharjee, and C.~Kommareddy, ``Scalable application layer
  multicast,'' in \emph{ACM SIGCOMM}, Aug 2002.

\bibitem{Banerjee2004}
S.~Banerjee, S.~Lee, R.~Braud, B.~Bhattacharjee, and A.~Srinivasan, ``Scalable
  resilient media streaming,'' in \emph{NOSSDAV'04}, June 2004.

\bibitem{Chu2000}
Y.~Chu, S.~G. Rao, S.~Seshan, and H.~Zuang, ``A case for end system
  multicast,'' in \emph{ACM SIGMETRICS}, June 2000.

\bibitem{Dominguez2007}
\BIBentryALTinterwordspacing
C.~Dominguez, H.~Ahlehagh, and C.~U. Pederson, ``Combining alm and customized
  codecs in improving video streaming,'' Tech. Rep., Sep 2007. [Online].
  Available: \url{http://web.mit.edu/carlosdc/www/report.pdf}
\BIBentrySTDinterwordspacing

\bibitem{Cohen2002}
E.~Cohen and S.~Shenker, ``Replication strategies in unstructured peer-to-peer
  networks,'' \emph{ACM SIGCOMM Computer Communication Review}, vol.~32, no.~4,
  2002.

\bibitem{Lam2004}
S.~S. Lam and H.~Liu, ``Failure recovery for structured p2p networks: protocol
  design and performance evaluation,'' in \emph{ACM SIGMETRICS '04/Performance
  '04}, June 2004.

\bibitem{Zhao2004}
B.~Y. Zhao, L.~Huang, J.~Stribling, S.~C. Rhea, A.~D. Joseph, and J.~D.
  Kubiatowicz, ``Tapestry: A resilient global-scale overlay for service
  deployment,'' \emph{IEEE Journal on Selected Areas in Communications},
  vol.~22, no.~1, 2004.

\bibitem{Rodrigues2005}
R.~Rodrigues and B.~Liskov, ``High availability in dhts: Erasure coding vs.
  replication,'' in \emph{International Workshop on Peer-to-Peer Systems
  IPTPS'05}, Feb 2005.

\bibitem{Weatherspoon2002}
H.~Weatherspoon and J.~Kubiatowicz, ``Erasure coding vs. replication: A
  quantitative comparison,'' in \emph{International Workshop on Peer-to-Peer
  Systems IPTPS'02}, Mar 2002.

\bibitem{Feldman2004}
M.~Feldman, C.~Papadimitriou, I.~Stoica, and J.~Chuang, ``Free-riding and
  white-washing in peer-to-peer systems,'' in \emph{ACM SIGCOMM workshop on
  Practice and Theory of Incentives and Game Theory in Networked Systems},
  2004.

\bibitem{Ma2004}
R.~T.~B. Ma, S.~M. Lee, J.~C.~S. Lui, and D.~K.~Y. Yau, ``An incentive
  mechanism for p2p networks,'' in \emph{Proceedings of the 24th International
  Conference on Distributed Computing Systems (ICDCS'04)}.\hskip 1em plus 0.5em
  minus 0.4em\relax IEEE, 2004.
\newpage
\bibitem{Wierzbicki2005}
A.~Wierzbicki, ``Peer-to-peer direct sales,'' in \emph{Proceedings of the Fifth
  IEEE International Conference on Peer-to-Peer Computing (P2P'05)}, 2005.

\bibitem{Wongrujira2005a}
K.~Wongrujira, T.~Hsin-Ting, and A.~Seneviratne, ``Incentive service model for
  p2p,'' in \emph{Computer Systems and Applications}, 2005.

\bibitem{Feldman2005a}
M.~Feldman and J.~Chuang, ``Overcoming free-riding behavior in peer-to-peer
  systems,'' \emph{ACM SIGecom Exchanges}, vol.~5, no.~4, p. 41–50, 2005.

\bibitem{Camerer2003}
C.~F. Camerer, \emph{Behavioral Game Theory}.\hskip 1em plus 0.5em minus
  0.4em\relax Princeton University Press, 2003.

\bibitem{Golle2001}
P.~Golle, K.~Leyton-Brown, I.~Mironov, and M.~Lillibridge, ``Incentives for
  sharing in peer-to-peer networks,'' in \emph{Proceedings of the 3rd ACM
  conference on Electronic Commerce}, October 2001.

\bibitem{Andrade2005}
N.~Andrade, M.~Mowbray, A.~Lima, G.~Wagner, and M.~Ripeanu, ``Influences on
  cooperation in bittorrent communities,'' in \emph{ACM SIGCOMM workshop on the
  Economics of Peer-to-Peer Systems}, 2005.

\bibitem{Cohen2003}
B.~Cohen, ``Incentives build robustness in bittorrent,'' in \emph{Workshop on
  Economics of Peer-to-Peer Systems}, 2003.

\bibitem{Huang2006a}
G.~Huang, P.~Hong, and J.~Li, ``Optimal server selection and pricing in
  receiver-driven p2p streaming systems,'' in \emph{ICON}, 2006.

\bibitem{Gupta2004}
R.~Gupta and A.~K. Somani, ``A pricing strategy for incentivizing selfish nodes
  to share resources in peer-to-peer (p2p) networks,'' in \emph{ICON}, 2004.

\bibitem{Moscibroda2006}
T.~Moscibroda, S.~Schmid, and R.~Wattenhofer, ``On the topologies formed by
  selfish peers,'' in \emph{ACM Symposium on Principles of Distributed
  Computing}, 2006, pp. 133--142.

\bibitem{Phillips2005}
R.~Phillips, \emph{Pricing and Revenue Optimization}.\hskip 1em plus 0.5em
  minus 0.4em\relax Palo Alto, CA: Stanford University Press, 2005.

\bibitem{Talluri2004}
K.~Talluri and G.~J. van Ryzin, \emph{The Theory and Practice of Revenue
  Management}.\hskip 1em plus 0.5em minus 0.4em\relax Dordrecht: Kluwer
  Academic, 2004.

\bibitem{Bitran2003}
G.~Bitran and R.~Caldentey, ``An overview of pricing models for revenue
  management,'' \emph{Manufacturing and Service Operations Management}, vol.~5,
  no.~3, p. 203–229, 2003.

\bibitem{Elmaghraby2003}
W.~J. Elmaghraby and P.~Keskinocak, ``Dynamic pricing: research overview,
  current practices, and future directions,'' \emph{Management Science},
  vol.~49, no.~10, p. 1287–1309, 2003.

\bibitem{Gallego1997}
G.~Gallego and G.~J. van Ryzin, ``A multi-product dynamic pricing problem and
  its applications to network yield management,'' \emph{Operations Research},
  vol.~45, no.~1, p. 24–41, 1997.

\bibitem{Kleywegt2001}
\BIBentryALTinterwordspacing
A.~J. Kleywegt, ``An optimal control problem of dynamic pricing,'' Georgia Tech
  Working Paper, Tech. Rep., 2001. [Online]. Available:
  \url{http://www.isye.gatech.edu/~anton}
\BIBentrySTDinterwordspacing

\bibitem{Maglaras2004}
\BIBentryALTinterwordspacing
C.~Maglaras and J.~Meissner, ``Dynamic pricing strategies for multi-product
  revenue management problems,'' Columbia University, Tech. Rep., 2004.
  [Online]. Available: \url{http://www.meiss.com}
\BIBentrySTDinterwordspacing

\bibitem{Paschalidis2000}
I.~C. Paschalidis and J.~N. Tstsiklis, ``Congestion dependent pricing of
  network services,'' \emph{IEEE/ACM Transactions on Networking}, vol.~8, p.
  171–184, 2000.

\bibitem{Yeung2006}
M.~K.~H. Yeung and Y.-K. Kwok, ``On maximizing reveue for client-server based
  wireless data access in the presence of peer-to-peer sharing,'' in \emph{The
  17th Annual IEEE International Symposium on Personal, Indoor and Mobile Radio
  Communications (PIMRC'06)}, 2006.

\bibitem{Zghaibeh2006}
M.~Zghaibeh and F.~C. Harmantzis, ``Lottery-based pricing scheme for peer to
  peer networks,'' in \emph{IEEE ICC 2006 proceedings}, 2006.

\end{thebibliography}
\end{document}